\documentclass[journal]{IEEEtran}
\usepackage{eso-pic}
\usepackage[utf8]{inputenc}
\usepackage{times,amsmath,epsfig,amssymb,cite,bm,float,epstopdf,graphicx,graphics,amsthm}
\usepackage{authblk}
\usepackage{balance,multicol}
\usepackage{color}
\usepackage{pdfsync}
\usepackage{ifthen}
\usepackage{cool}
\usepackage{makeidx}
\usepackage{verbatim}
\usepackage{algorithm, algorithmic, enumerate}
\usepackage{stfloats}
\usepackage{pgfplots}
\usepackage{algorithm}
\usepackage{algorithmic}
\usepackage{amsmath}

\usepackage{amsfonts} 
\usepackage{multirow,array} 
\usepackage{booktabs}
\usepackage{tikz}
\usepackage{enumitem}
\usetikzlibrary{automata,arrows,positioning,calc}
\usepackage{lipsum}
\usepackage{threeparttable} 
\usepackage{blindtext}
\PassOptionsToPackage{hyphens}{url}
\usepackage[bookmarks=false]{hyperref}
\usepackage{gensymb}
\usepackage{chemformula}
\usepackage{tabu} 
\usepackage{tabularx}
\usepackage{makecell}
\usepackage{cite}
\usepackage{siunitx}
\usepackage{colortbl}
\usepackage{subcaption}

\def\isfinal{0}
\newcommand{\wipcomment}[2]{\ifnum\isfinal=0{\color{#1}{#2}}\else{#2}\fi}

\title{Lightweight Latency Prediction Scheme for Edge Applications: A Rational Modelling Approach}

\author{ 
 \IEEEauthorblockA{Mohan Liyanage, Eldiyar Zhantileuov, Ali Kadhum Idrees,  Rolf Schuster            ~~~~~~~~~~~~~~~~~~~~~~~~~~~~~~~~~~~~~~
University of Applied Sciences and Arts, Computer Science Department, Dortmund, Germany. eldiyar.zhantileuov@fh-dortmund.de}
 }

\begin{document}

\makeatletter
\AddToShipoutPicture*{%
  \put(55,770){%
    \parbox[t]{\textwidth}{%
      \raggedright
      \fontsize{8}{10}\selectfont
      Presented at the ICCS 2025 — 5th International Conference on Computer Systems,  Xi'an, China. %
    }%
  }%
}
\makeatother

\maketitle

\begin{abstract}

Accurately predicting end-to-end network latency is essential for enabling reliable task offloading in real-time edge computing applications. This paper introduces a lightweight latency prediction scheme based on rational modelling that uses features such as frame size, arrival rate, and link utilization, eliminating the need for intrusive active probing. The model achieves state-of-the-art prediction accuracy through extensive experiments and 5-fold cross-validation  (MAE = 0.0115, R$^2$ = 0.9847) with competitive inference time, offering a substantial trade-off between precision and efficiency compared to traditional regressors and neural networks.

\end{abstract}

\begin{IEEEkeywords}
Edge computing, Network latency prediction, Quality of Service (QoS), Rational-exponential model
\end{IEEEkeywords}

\section{Introduction}

Edge computing has emerged as a critical paradigm for latency-sensitive applications that require real-time processing close to data sources. In scenarios such as intelligent transportation systems, augmented reality, and autonomous driving, the ability to process and respond to data with minimal end-to-end (E2E) delay is central to maintaining high Quality of Service (QoS). Among these, real-time video analytics for connected vehicles—such as traffic sign detection—demands ultra-low latency to ensure both functionality and safety.

A key determinant of QoS in edge environments is the ability to accurately estimate network latency and make timely offloading decisions. Conventional approaches often rely on machine learning (ML) models or analytical heuristics to predict delay, but many of these suffer from high inference latency or limited generalizability. This makes them suboptimal for use cases where the prediction itself must be computed quickly to preserve responsiveness. For instance, deploying deep learning models for delay estimation in a real-time edge pipeline can introduce delays that negate their predictive utility.

We propose a lightweight, numerically efficient framework that combines real-time delay prediction with adaptive offloading decision logic to address this challenge. Our system monitors traffic conditions across multiple network segments and uses a rational function-based model to estimate E2E delay with high accuracy and minimal computational overhead. 
We conducted extensive real-world experiments to validate our framework using a campus-deployed private 5G network infrastructure. This environment includes high-performance edge servers co-located with the 5G core, enabling realistic testing of multi-tier offloading scenarios. The experimental results confirm that our QoS-aware offloading mechanism remains robust and effective even under highly variable network conditions, offering a practical and scalable solution for real-time edge applications.

\section{Related work} \label{sec:relwo}
Accurate prediction of network-level latency is essential to ensure the reliability of edge computing applications, particularly in scenarios that require real-time processing and low-latency performance. In recent years, various methods have been proposed, including analytical models, machine learning (ML)-based predictors, and hybrid techniques that integrate both approaches. This section comprehensively reviews these cutting-edge methods, focusing on environments involving single-edge nodes and predicting real-time Round Trip Time (RTT) or transmission delays.
\subsection{Analytical Approaches}
Analytical models rely on queueing theory, network calculus, or stochastic geometry to derive expected latency based on known parameters such as arrival rates, service times, or spatial distributions.  These models are both lightweight and interpretable. For instance, M/M/1 queues and their extensions have been employed to estimate average and worst-case delays~\cite{bertsekas2021data,miao2024performance}. One of the seminal formulations in this area is the Kleinrock function~\cite{1574231874180177024}, which approximates the average delay under the assumption of Poisson arrivals. Network calculus, in particular, provides deterministic bounds for worst-case delays in systems with burst traffic~\cite{le2002network}. Other works  such as  Fortz and Thorup~\cite{fortz2000internet} and Altman et al.~\cite{altman2002competitive} extended these models to more realistic traffic scenarios.

 
\subsection{Machine Learning-Based Models}
With the rise of software-defined networking and the availability of extensive telemetry data, recent research has increasingly concentrated on data-driven approaches to enhance the modeling of Quality of Service (QoS) metrics. In their study, Krasniqi et al.~\cite{krasniqi2020end} proposed a machine learning-based model aimed at predicting end-to-end delay through traffic matrix sampling. Their simulation framework, developed using NS-3, generated realistic datasets that reflected variations in traffic intensity, link capacities, and propagation delays. 
In~\cite{mestres2018understanding}, the authors used a fully connected feed-forward neural network to predict average network delay from traffic matrix inputs. They aimed to explore how traffic characteristics, like intensity, affect neural network performance. 
Deep-Q~\cite{xiao2018deep} is a sophisticated data-driven framework designed to model the distribution of Quality of Service (QoS) metrics within a network based on observed traffic conditions. It utilizes a Deep Generative Network (DGN) architecture that integrates Long Short-Term Memory (LSTM) networks with a Variational Autoencoder (VAE). 
Utilizing  Graph Neural Networks (GNNs), RouteNet~\cite{rusek2019unveiling} demonstrates commendable generalization but also presents challenges related to the training and deployment of GNNs. The framework predicts delay and jitter across diverse network topologies and routing configurations.

\subsection{Hybrid and Adaptive Models}
In latency-sensitive systems, hybrid models integrating deep learning with statistical or clustering techniques have demonstrated significant potential in enhancing prediction accuracy while maintaining interpretability. For example, the research presented in~\cite{zhang2020latency} introduces a dual-component latency prediction model specifically designed for vehicular applications. The first component utilizes a Long Short-Term Memory (LSTM) network to capture temporal trends in latency, with an initial application of k-medoids clustering to group similar latency patterns, thereby improving prediction performance. The second component employs a statistical method based on Epanechnikov kernel smoothing and moving average functions to refine the predictions further. 
Another noteworthy example of a hybrid machine learning approach is ViCrypt, detailed in~\cite{seufert2019features}, which focuses on real-time Quality of Experience (QoE) monitoring for encrypted YouTube video streams. Due to end-to-end encryption limiting access to payload information, ViCrypt exclusively relies on traffic patterns of encrypted data and derives lightweight statistical features on a stream-by-stream basis. 

Additionally, another area of research focuses on latency optimization in service function chaining (SFC) and the placement of Virtual Network Functions (VNF). As discussed in~\cite{gouareb2018delay}, the authors explore how VNFs and multipath flow routing placement affect accumulated network delays. VNFs, hosted within virtual machines, must be sequenced appropriately to fulfil service requests, making the order, placement, and routing of functions critical to latency and resource efficiency.

\section{Scenario: QoS-Aware Server Selection in Edge Video Analytics }
\label{sec:qos_offloading_Scenario}
This section presents the use case scenario which is implemented over a campus 5G network equipped with high-performance edge servers and multi-segment monitoring framework.

\begin{figure}[htbp]
    \centering
    \includegraphics[width=0.9\linewidth]{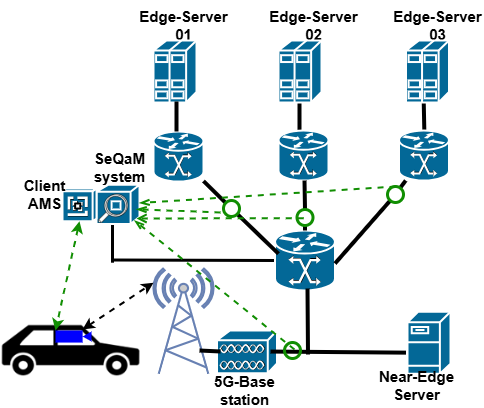}
    \caption{Network model for experimental setup}
    \label{fig:scenario}
\end{figure}
In latency-sensitive edge computing applications, such as real-time video analytics for intelligent vehicles, ensuring timely processing is paramount. The quality of service (QoS), particularly in terms of end-to-end (E2E) delay, directly influences the client application's operational viability. Figure~\ref{fig:scenario} illustrates a representative use case, where a vehicle-mounted camera captures video streams (e.g., for traffic sign recognition) and offloads the frames to a selected computing node for processing.
In this system, three candidate locations are available for video frame processing: the on-board vehicle computer, a Near-Edge server co-located with the 5G base station, and one of several high-performance Edge servers located deeper (e.g., two hops away) in the network infrastructure. Each of these offloading targets presents a different trade-off between computational power and communication delay.
The vehicle’s local computer offers the lowest E2E latency by eliminating network transmission altogether. However, due to its limited computational capacity, it is only suitable for lightweight models or temporary fallback scenarios. This option is typically activated when the 5G access delay exceeds a predefined threshold, signaling that edge offloading is currently infeasible.
The Near-Edge server, situated at the 5G base station, provides moderate processing capability but benefits from reduced network latency. It serves as an intermediate fallback when the performance of remote edge servers degrades due to congestion, yet the 5G uplink remains within acceptable delay bounds.

The remote high-performance Edge servers offer superior processing speed and can handle complex deep learning inference tasks. Under nominal network conditions, this tier serves as the default destination for data offloading. However, since these servers are accessible over longer or more congested paths, their E2E delay may fluctuate significantly. To mitigate this, the system includes multiple redundant servers with similar capabilities, allowing the client application to dynamically switch between them based on real-time network metrics. 
A Service Quality Manager for Edge Computing (SeQaM)~\cite{burbano2024end} has been implemented to facilitate proactive and intelligent server selection. SeQaM serves as an Edge monitoring and management platform that offers distributed observability, adaptive data collection, real-time analytics, rapid feedback mechanisms, and the capability to generate controlled events and experimental scenarios. 
The predicted delay values are then transmitted to a centralized Client Application Management System (CAMS), which aggregates the data and makes informed decisions on behalf of the client.

\section{Methodology}
\label{sec:Methodology}
This section presents the methodology used to design a lightweight, reliable delay prediction model and integrate it into a real-time offloading decision engine.
\subsection{Network Monitoring and Delay Estimation}
To support delay-aware offloading decisions, four delay segments are continuously monitored within the system:

\begin{itemize}
    \item $D_{\text{5G}}$: Delay on the wireless 5G uplink.
    \item $D_{E1}, D_{E2}, D_{E3}$: Delay on the respective paths to each of the three high-performance edge servers.
\end{itemize}
Dedicated monitoring stations are strategically placed along each segment to measure real-time network conditions. These stations collect metrics (such as \textit{link utilization, throughput} and \textit{packet arrival rates}), which are then used to estimate the current delay as mentioned in the previous section.
\subsection{Delay-Aware Decision Logic}
\label{sec:delay_logic}
We define the total expected end-to-end delay for each candidate offloading target as follows:
\begin{itemize}
    \item \textbf{Local processing (on-vehicle)}:
    \begin{equation}
        T_{\text{local}} = D_{\text{proc}}^{\text{local}}
    \end{equation}

    \item \textbf{Near-edge server (at base station)}:
    \begin{equation}
        T_{\text{near}} = D_{5G} + D_{\text{proc}}^{\text{near}}
    \end{equation}

    \item \textbf{High-performance edge servers (for } $i \in \{1,2,3\}$\textbf{)}:
    \begin{equation}
        T_{\text{edge}_i} = D_{5G} + D_{E_i} + D_{\text{proc}}^{\text{edge}_i}
    \end{equation}
\end{itemize}
Since our primary focus is on characterizing network-induced delay variations, we treat the server-side processing delays $D_{\text{proc}}^{(\cdot)}$ as either negligible or constant. These components are not explicitly predicted but are included symbolically for completeness.
To summarize, Table~\ref{tab:delay_components} provides a consolidated view of all delay components used in our offloading logic.

\begin{table*}[htbp]
\centering
\caption{End-to-End Delay Components for Offloading Destinations }
\normalsize
\begin{tabular}{l l l}
\hline
\textbf{Destination} & \textbf{Total Delay Expression} & \textbf{Description} \\
\hline
\textbf{Local (L-Comp)} &
$T_{\text{local}} = D_{\text{proc}}^{\text{local}}$ &
Execution on vehicle with no network delay \\
\textbf{Near-Edge Server} &
$T_{\text{near}} = D_{5G} + D_{\text{proc}}^{\text{near}}$ &
5G uplink delay + processing at near-edge (constant) \\
\textbf{Edge Server 1} &
$T_{\text{edge}_1} = D_{5G} + D_{E1} + D_{\text{proc}}^{\text{edge}_1}$ &
5G delay + path to Edge 1 + processing delay (constant) \\
\textbf{Edge Server 2} &
$T_{\text{edge}_2} = D_{5G} + D_{E2} + D_{\text{proc}}^{\text{edge}_2}$ &
5G delay + path to Edge 2 + processing delay (constant) \\
\textbf{Edge Server 3} &
$T_{\text{edge}_3} = D_{5G} + D_{E3} + D_{\text{proc}}^{\text{edge}_3}$ &
5G delay + path to Edge 3 + processing delay (constant) \\
\hline
\end{tabular}
\label{tab:delay_components}
\end{table*}

\subsection{Rational-Exponential Model for Delay Prediction}

To capture compounding nonlinear effects—such as sudden spikes in latency due to network saturation—we extend the rational delay model by incorporating an exponential modulation. The predicted delay $\hat{y}$ is now defined by Equation~\ref{eq:rational_exp_delay_model}:

\begin{equation}
\hat{y}(x_1, x_2, x_3) = 
\left(
\frac{a_1 x_1 + a_2 x_2 + a_3 x_3}{1 + b_1 x_1 + b_2 x_2 + b_3 x_3 + c}
\right)
\cdot \exp(d \cdot x_3)
\label{eq:rational_exp_delay_model}
\end{equation}
where:
\begin{itemize}
    \item \( x_1 \) = Client\_Frame\_Size (bytes)
    \item \( x_2 \) = Network utilization (\%)
    \item \( x_3 \) = Arrival\_rate\_All (packets/s)
    \item \( a_1, a_2, a_3, b_1, b_2, b_3, c, d \) are coefficients estimated from training data
\end{itemize}
This formulation enhances the classical rational model (Equation~\ref{eq:general_rational}) by introducing an exponential sensitivity term with respect to arrival rate. 
\begin{equation}
Y = \frac{a_1 X^{a_2}}{1 + a_3 X^{a_4}}
\label{eq:general_rational}
\end{equation}
Multivariable extensions of rational functions have demonstrated remarkable effectiveness in modelling complex systems' intricate behaviours while maintaining a streamlined number of parameters~\cite{ponton1993use}. 
Such hybrid models have been found effective in representing sharp transitions and high-load instability in real-time systems~\cite{motulsky1987fitting,draper1998applied}.
The decision logic that incorporates both predicted delay and node reliability is summarized in Algorithm~\ref{alg:offloading_reliability}. Each candidate node's total delay is normalized, and a composite score is computed by balancing delay and reliability.

\begin{algorithm}[htbp]
\caption{Delay and Reliability-Aware node Selection}
\label{alg:offloading_reliability}
\begin{algorithmic}[1]
\REQUIRE $D_{5G}$, $D_E[1..3]$, $D_{\text{proc}}^{\text{local}}$, $D_{\text{proc}}^{\text{near}}$, $D_{\text{proc}}^{\text{edge}}[1..3]$, $\delta_{\text{max}}$, reliability scores $R_j$
\ENSURE Selected offloading target

\IF{$D_{5G} > \delta_{\text{max}}$}
    \STATE \textbf{return} (\texttt{LOCAL}, $D_{\text{proc}}^{\text{local}}$)
\ENDIF

\STATE Compute $T_{\text{local}}, T_{\text{near}}, T_{\text{edge}_1}, T_{\text{edge}_2}, T_{\text{edge}_3}$
\STATE $C \gets$ all total delays $\{T_j\}$

\STATE $T_{\max} \gets \max_{j} T_j$
\STATE $\alpha \in [0, 1]$ \COMMENT{Trade-off weight}

\FORALL{$j \in C$}
    \STATE $S_j \gets \alpha \cdot \frac{T_j}{T_{\max}} + (1 - \alpha) \cdot (1 - R_j)$
\ENDFOR

\STATE $j^* \gets \arg\min_j S_j$
\STATE \textbf{return} $(j^*, T_{j^*})$
\end{algorithmic}
\end{algorithm}

\section{Experimental Results and Discussion} \label{sec:evaluation}

\subsection{Experimental Setup}

The evaluation of our proposed end-to-end latency prediction and offloading framework was conducted on a private 5G campus network. This real-world deployment includes:
\begin{itemize}
    \item A client capturing video frames and uploading to the server.
    \item A 5G radio access network connected to:
    \begin{itemize}
        \item A {Near-Edge server} co-located with the base station.
        \item Three {High-Performance Edge servers} (Edge1–Edge3) located across the wired network at increasing path lengths.
    \end{itemize}
    \item The {SeQaM} platform responsible for capturing link-level metrics every 100 milliseconds across all segments: client, the 5G uplink and the links to each server.
\end{itemize}
The collected dataset includes over 5000 samples across varying network conditions and frame sizes. Ground truth delay measurements were timestamped from the moment a video frame was transmitted by the client to the moment a response was received.
\subsection{Feature Engineering for Edge Delay Modeling}
Effective feature engineering is essential for developing predictive models that are both accurate and interpretable in highly dynamic edge environments. In our campus-scale scenario involving video frame uploads from vehicles to edge servers (as detailed in Section~\ref{sec:qos_offloading_Scenario}), delay behavior is influenced by multiple concurrent variables, including image data volume, network link load, and overall traffic.
We focus on constructing features that improve the performance of our rational function-based delay prediction model for edge network traffic. Due to the complexity and non-linearity inherent in network behaviors, careful selection and transformation of features are important for accurately capturing the underlying dynamics of delay.
\subsection{Initial Feature Set}
The raw dataset collected from the 5G deployment includes the following features:
\begin{itemize}
\item {Client\_Frame\_Size}: Size of the video frame transmitted (in bytes).
\item {Arrival\_rate\_Cl}: Observation of the arrival rate of packets from the client interface.
\item {Arrival\_rate\_All}: Aggregate arrival rate from all connected devices on the segment.
\item {Utilization}: Total utilization of monitored segments.
\end{itemize}

{Note:} Throughout this study, 
\textit{Utilization} strictly refers to the utilization of the network link (not CPU usage).

The prediction target is {Delay} (end-to-end), measured in seconds from frame upload to edge response reception.
Certain~features~like,~\texttt{Client\_Frame\_Size},~\texttt{Arrival\_ rate\_Cl}, \texttt{Arrival\_rate\_All} have significantly larger magnitudes compared to others. This scale disparity can hinder optimization during curve fitting. Therefore, we apply manual rescaling which enhances numerical stability while maintaining the original distribution. 
\subsection{Feature Selection via Correlation and Sensitivity Analysis}
Pearson's correlation coefficients and mean absolute sensitivity were used to evaluate feature redundancy and model impact. \texttt{Arrival\_rate\_Cl} and \texttt{Client\_Frame\_Size} exhibited a high correlation (0.98), indicating redundancy. To reduce multicollinearity, \texttt{Arrival\_rate\_Cl} was removed.~After filtering, the following features were retained: 1) \texttt{Client\_Frame\_Size} (rescaled), 2) \texttt{Arrival\_rate\_All} (rescaled), 3) \texttt{Utilization.}

\subsection{Justification for the Rational-Exponential Model}
\label{sec:methodology_rational_exp_justification}

To determine the most suitable functional form for modeling delay, we compared multiple candidate models—including linear regression, polynomial regression, sigmoid-based formulations, a multilayer perceptron (MLP), and two rational function families. Each model was evaluated in terms of prediction accuracy (MAE, MSE, $R^2$) and real-time inference efficiency.

\begin{table*}[h]
\centering
\caption{Comparison of Delay Prediction Models}
\small
\begin{tabular}{lcccccc}
\toprule
\textbf{Model} & \textbf{MAE} & \textbf{MSE} & \textbf{$R^2$} & \textbf{Avg Time (ms)} & \textbf{Min (ms)} & \textbf{Max (ms)} \\
\midrule
Rational & 0.01273 & 0.00032 & 0.9843 & 0.0215 & 0.0167 & 0.1411 \\
Rational-Exponential & \textbf{0.01215} & \textbf{0.00031} & \textbf{0.9848} & \textbf{0.0215} & \textbf{0.0188} & \textbf{0.0370} \\
Polynomial (2nd Order) & 0.01359 & 0.00035 & 0.9839 & 0.0464 & 0.0238 & 0.1144 \\
Linear Regression & 0.01542 & 0.00039 & 0.9821 & 0.0186 & 0.0141 & 0.0563 \\
Neural Network (MLP) & 0.01493 & 0.00037 & 0.9831 & 0.1264 & 0.0987 & 0.1982 \\
Sigmoid & 0.24842 & 0.08228 & -3.0555 & 0.0204 & 0.0143 & 0.1235 \\
\bottomrule
\end{tabular}
\label{tab:model_comparison}
\end{table*}

Among all candidates, the {Rational-Exponential model} achieved the best overall trade-off. As exhibited in Table~\ref{tab:model_comparison}, the lowest mean absolute error (MAE = 0.0121), smallest mean squared error (MSE = 0.00031), and highest coefficient of determination ($R^2 = 0.9848$). Additionally, it maintained a low average inference latency of 0.0215 ms over 100 test samples, making it highly suitable for edge deployments. In contrast, traditional neural networks such as MLPs, while expressive, incurred higher computational costs and demonstrated less stability across the dataset. Similarly, sigmoid-based models underperformed in both accuracy ($R^2 = -3.06$) and numerical stability due to exponential saturation effects and coefficient overfitting.

\begin{figure}[h]
\centering
\includegraphics[width=1\linewidth]{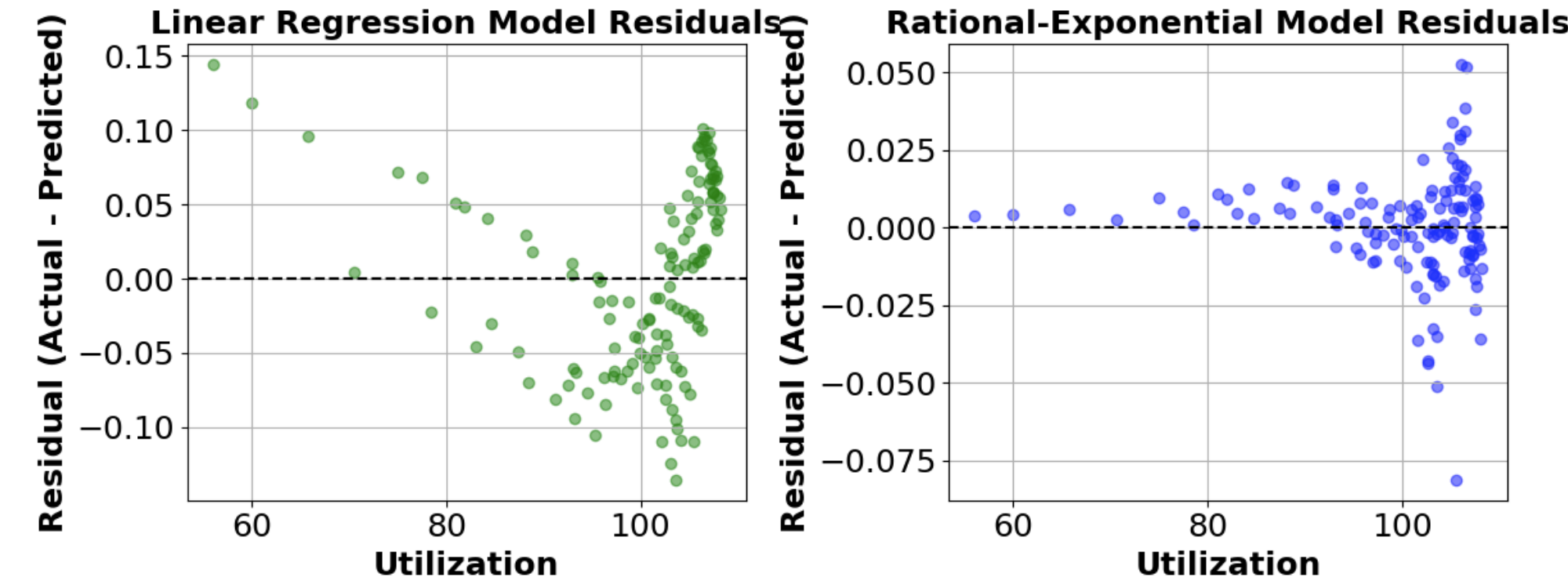}
\caption{Residual analysis of the Linear Regression (left) and Rational-Exponential (right) models plotted against \texttt{Utilization} }
\label{fig:residuals_utilization}
\end{figure}

The residual plots in Figure~\ref{fig:residuals_utilization} reveal critical differences in model behavior under varying network load conditions. The Linear Regression model (left plot) shows a clear bias in the residuals as \texttt{Utilization} increases. Specifically, in the high utilization range (above 100), the residuals exhibit a wider spread and tend to skew positive, indicating that the model tends to underestimate delays under congestion. Such behavior is problematic for real-time decision-making, where accurate predictions during saturation are essential.~
In contrast, the Rational-Exponential model (right plot) demonstrates residuals that are tightly clustered around zero across the entire utilization range. Even under high load, the residuals remain symmetrically distributed with minimal bias, indicating better generalization and stability. This consistency suggests that the Rational-Exponential model effectively captures the nonlinear escalation of delay as the network becomes saturated. These results highlight the Rational-Exponential formulation as a more robust and reliable choice for delay prediction in real-time offloading scenarios.

To benchmark the generalization performance of different delay prediction models, we conducted 5-fold cross-validation across all candidates. As summarized in Table~\ref{tab:cv_results}, the proposed \textit{Rational-Exponential} model achieved the best overall results, with the lowest mean absolute error (MAE = 0.0115~$\pm$~0.0029) and highest coefficient of determination (R$^2$ = 0.9847~$\pm$~0.0094), indicating excellent predictive accuracy and stability. Polynomial regression offered comparable accuracy, while linear regression underperformed due to its inability to model nonlinear trends. The neural network model (MLP) exhibited significant overfitting, leading to negative R$^2$ scores and high variance.
\begin{table*}[htbp]
\centering
\caption{5-Fold Cross-Validation Results for All Models}
\small
\begin{tabular}{lcccccc}
\toprule
\textbf{Model} & \textbf{MAE} & \textbf{MAE Std} & \textbf{MSE} & \textbf{MSE Std} & \textbf{R$^2$} & \textbf{R$^2$ Std} \\
\midrule
Rational & 0.0127 & 0.0029 & 0.00032 & 0.00017 & 0.9843 & 0.0094 \\
Rational-Exponential & \textbf{0.0115} & \textbf{0.0029} & \textbf{0.00029} & \textbf{0.00017} & \textbf{0.9847} & \textbf{0.0094} \\
Polynomial (2nd Order) & 0.0146 & 0.0018 & 0.00042 & 0.00014 & 0.9784 & 0.0066 \\
Linear Regression & 0.0553 & 0.0028 & 0.00409 & 0.00031 & 0.7873 & 0.0359 \\
Neural Network (MLP) & 0.1709 & 0.0266 & 0.0434 & 0.0112 & –1.2893 & 0.8231 \\
\bottomrule
\end{tabular}
\label{tab:cv_results}
\end{table*}

\section{Summary and Conclusions} \label{sec:conclusions}

This paper presented a lightweight framework for end-to-end latency prediction in real-time edge computing applications. We developed a highly accurate yet computationally efficient prediction model by leveraging readily available network features and modelling delay using Rational-Exponential functions. Compared to classical regression and neural network baselines, the Rational-Exponential model consistently achieved excellent performance, with an R$^2$ of 0.9847 and MAE as low as 0.0115, as validated through 5-fold cross-validation and residual analysis. The offloading evaluations confirmed that the Rational-Exponential model yielded the highest decision accuracy across multiple candidate paths.



\balance{}
\bibliographystyle{IEEEtran}
\bibliography{Latency}

\end{document}